\documentclass[%
reprint,
superscriptaddress,
amsmath,amssymb,
aps,
prb,
]{revtex4-2}

\usepackage{booktabs}
\usepackage{multirow}
\usepackage{color}
\usepackage{graphicx}
\usepackage{dcolumn}
\usepackage{bm}
\usepackage{hyperref}
\usepackage[mathlines]{lineno}
\usepackage{tabularx}

\begin{document}
	
	\title{Robust and tunable Floquet altermagnets in sliding A-type antiferromagnetic bilayers}
	
	\author{Zhe Li}%
	\email{lizhe21@iphy.ac.cn}
	\affiliation{%
		Beijing National Laboratory for Condensed Matter Physics, and Institute of Physics, Chinese Academy of Sciences, Beijing 100190, China
	}%

	\author{Lijuan Li}
	\affiliation{%
		Centre for Quantum Physics, Key Laboratory of Advanced Optoelectronic Quantum Architecture and Measurement (Ministry of Education), School of Physics, Beijing Institute of Technology, Beijing 100081, China
	}%

	\author{Mengxue Guan}
		\email{mxguan@bit.edu.cn}
	\affiliation{%
		Centre for Quantum Physics, Key Laboratory of Advanced Optoelectronic Quantum Architecture and Measurement (Ministry of Education), School of Physics, Beijing Institute of Technology, Beijing 100081, China
	}%

	\author{Sheng Meng}
	\affiliation{%
		Beijing National Laboratory for Condensed Matter Physics, and Institute of Physics, Chinese Academy of Sciences, Beijing 100190, China
	}%
	
	\affiliation{%
		University of Chinese Academy of Sciences, Beijing 100049, China
	}%
	
	\affiliation{%
		Songshan Lake Materials Laboratory, Dongguan, Guangdong 523808, China
	}%

	\date{\today}
	
	\begin{abstract}
		Arranging the stacking orders of A-type antiferromagnetic (A-AFM) bilayers offers a practical pathway to realizing two-dimensional altermagnets. Previous proposals, however, rely on stringent symmetry constraints, including layer groups and stacking registries. In this work, we demonstrate that circularly polarized light (CPL) irradiation breaks time-reversal symmetry, enabling altermagnetism without these restrictions. Our comprehensive symmetry analysis reveals that inversion-symmetric A-AFM bilayer building blocks can host altermagnetism that is robust against stacking slides and variations in illumination direction, with rich and continuously tunable symmetry representations. Using bilayer \(\mathrm{MnBi_{2}Te_{4}}\) as a case study, we show that out-of-plane CPL stabilizes distinct $f$-wave or $p$-wave altermagnetic states in forward stacking, while reverse stacking manifests an even-odd parity conversion of altermagnetism based upon specific sliding configurations. These revealments broaden the scope of altermagnet design in inversion-symmetric A-AFM bilayers and establish CPL as a versatile control knob for engineering altermagnetic symmetries.
	\end{abstract}
	
	\maketitle


    
	Recently, the simplification and generalization of material design principle for altermagnets have emerged as one of the most active directions in condensed-matter reserach\cite{vsmejkal2022beyond,vsmejkal2022emerging,bai2024altermagnetism,song2025altermagnets,xun2025stacking,zhu2025sliding,chen2024enumeration,chen2025unconventional,gu2025ferroelectric,duan2025antiferroelectric}. Altermagnets are characterized by the absence of net magnetization while exhibiting symmetry-protected spin splitting in specific regions of the Brillouin zone (BZ), representing an unconventional class of magnetic order \cite{vsmejkal2022beyond,vsmejkal2022emerging,bai2024altermagnetism,song2025altermagnets,xun2025stacking,zhu2025sliding,gu2025ferroelectric,chen2024enumeration,chen2025unconventional,duan2025antiferroelectric}. Spin-space symmetry operators enforce spin degeneracy at specific points, lines, or planes in momentum space, forming boundaries between regions of opposite spin splitting. The geometry of these boundaries determines the spin-dependent Fermi surfaces and their associated planar or bulk $d$-wave, $g$-wave, or $i$-wave symmetries \cite{vsmejkal2022beyond,vsmejkal2022emerging,bai2024altermagnetism,song2025altermagnets,xun2025stacking}. To date, a variety of altermagnets have been theoretically proposed  or experimentally identified, including MnTe \cite{krempasky2024altermagnetic,lee2024broken,liu2024chiral}, CrSb \cite{ding2024large,zeng2024observation,yang2025three}, RuO$_2$ \cite{vsmejkal2023chiral,feng2022anomalous,bai2022observation,zhou2024crystal}, KV$_2$Se$_2$O \cite{jiang2025metallic,xu2025electronic}, Rb$_{1-\delta}$V$_2$Te$_2$O \cite{zhang2024crystal}, Nb(V)$_2$SeTeO \cite{xun2025stacking,xie2025piezovalley,feng2025type,zou2025floquet}, NiZrI$_6$ \cite{pan2024general,zeng2024bilayer,zhu2023multipiezo}, among others \cite{mazin2021prediction,bhowal2024ferroically,morano2025absence,guo2023spin,leiviska2024anisotropy,reichlova2024observation,badura2025observation,wang2022magneto}. Notably, altermagnets exhibit unique behaviors, including large anomalous Hall effect \cite{feng2022anomalous,sato2024altermagnetic,liu2025anomalous,leiviska2024anisotropy,reichlova2024observation}, anomalous Nernst effect \cite{badura2025observation,yi2025spin}, crystal thermal transport \cite{zhou2024crystal}, piezomagnetism \cite{zhu2023multipiezo,jiang2025strain}, piezoelectricity \cite{blonsky2015ab}, spin-splitter torque \cite{bai2022observation,gonzalez2021efficient}, the magneto-optical effect \cite{zhou2021crystal,wang2022magneto}, chiral magnons \cite{vsmejkal2023chiral}, and the type-II quantum spin Hall effect \cite{feng2025type}, and beyond \cite{bai2022observation,gonzalez2021efficient}.
	
	Among the various realization schemes, a direct strategy for constructing altermagnets relies on the deliberate stacking of A-type antiferromagnetic (A-AFM) bilayers with specific stacking registries, as prescribed by general stacking theory \cite{pan2024general,zeng2024bilayer,sun2024stacking,zhu2025sliding}. However, this approach imposes several stringent constraints. First, only 17 out of the 80 crystallographic layer groups intrinsically support altermagnetism, and only the symmetry operators $C_{2\alpha}$ or $S_{4z}$ are compatible with this magnetic order \cite{zeng2024bilayer}. Second, the emergence of altermagnetism within stacking theory is highly sensitive to particular stacking registries, including precise sliding configurations or twist angles \cite{pan2024general,zeng2024bilayer,sun2024stacking,zhu2025sliding,zhang2025altermagnetic}, which significantly limits experimental feasibility. Furthermore, the presence of inversion symmetry ($\left[C_2||P\right]$) or mirror symmetry with respect to the $x$-$y$ plane ($\left[C_2||M_z\right]$) suppresses altermagnetim \cite{pan2024general,zeng2024bilayer,huang2025light}. Notably, inversion symmetry is prevalent in many conventional materials, which substantially limits candidate platforms.
	
	Recently, Floquet engineering \cite{liu2023floquet,blanes2009magnus,bukov2015universal,li2025light,li2025laser,li2025light-plumbene,qiu2018ultrafast} using circularly or elliptically polarized light (CPL or EPL) has opened up alternative routes towards altermagnets \cite{huang2025light,li2025floquet,zhu2025floquet,liu2025light,luo2025spin,zhuang2025odd,ghorashi2025dynamical}. In contrast to altermagnets protected by time-reversal symmetry ($\left[C_2||T\right]$), which enforces the even-parity relationship $\varepsilon(\textbf{\textit{k}},\ s)=\varepsilon(-\textbf{\textit{k}},\ s)$,  CPL or EPL explictly breaks this symmetry \cite{huang2025light,li2025floquet,zhu2025floquet,liu2025light,luo2025spin,zhuang2025odd,ghorashi2025dynamical}. As a result, odd-parity altermagnetic states become accessible, and the constraints imposed by inversion $\left[C_2||P\right]$ or mirror symmetry $\left[C_2||M_z\right]$ are lifted. Consequently, altermagnetism can emerge in a broad class of conventional A-AFM bilayer systems previously considered symmetry-forbidden.
	
	In this work, based on a systematic symmetry analysis of spin-space group, we demonstrate that under the Floquet engineering with CPL, all inversion-symmetric A-AFM bilayers host odd-parity altermagnetism, independent of stacking sliding or irradiation angles. Such A-AFM bilayers can be generically constructed from arbitrary ferromagnetic (FM) monolayer via a combined  $C_{2y}M_y$ operation relative to a point slightly above the first layer, consistent with the $d$-electron counting rule \cite{xiao2020electron,li2020tunable,tang2023intrinsic}. The only exception arises from the symmetry operator $\left[C_2 || M_z\right]$, which suppresses altermagnetism exclusively under $z$-axis irradiation. Utilizing bilayer $\mathrm{MnBi_{2}Te_{4}}$ (BL-MBT) as a benchmark, we find that forward-stacking BL-MBT exhibits $f$-wave altermagnetism under CPL along the $z$-axis, transitioning to $p$-wave altermagnetism with stacking sliding or varying incident angles. In contrast, reversed stacking of BL-MBT loses inversion symmetry but enables CPL-engineered $d$-$p$ conversion of altermagnetism at specific stacking configurations. Our findings underscore the broad potential of altermagnetism in inversion-symmetric A-AFM bilayers, providing a simple and effective method for designing altermagnets.
	

	\begin{figure}
		\centering
		\includegraphics[width=0.8\linewidth]{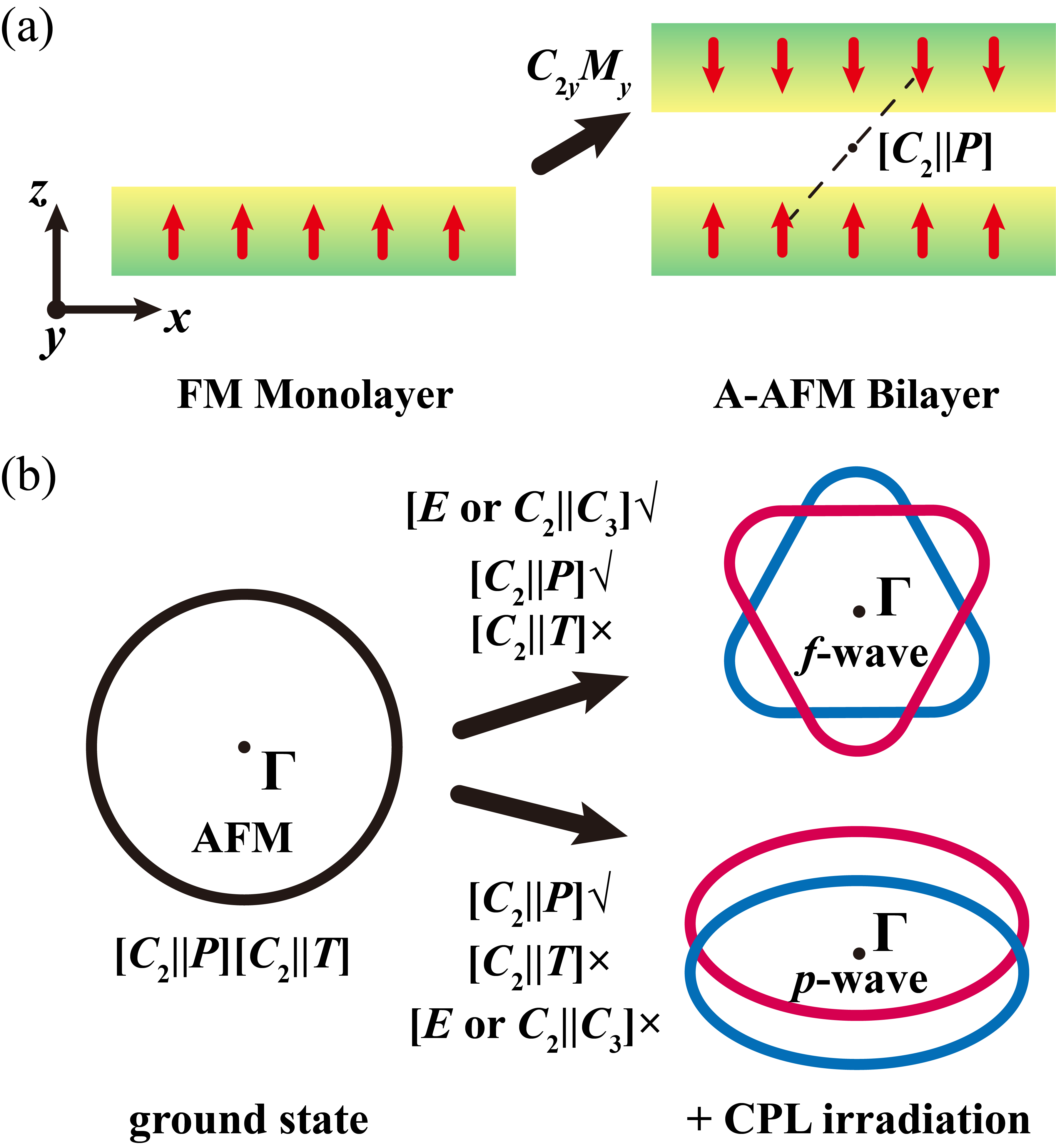}
		\caption{\textbf{Schematic of establishing Floquet altermagnets.} (a) Constructing strategy of an A-AFM bilayer with $\left[\textit{C}_\mathbf{2}||\textit{P}\right]$ from a FM monolayer. The yellow and green regions represent the top and bottom parts of the FM monolayer, respectively, while the red arrows indicate the magnetic moments. (b) Illustrations of light-induced spin-splitting patterns in $f$- or $p$-wave altermagnets, where the red and blue loops represent spin-up and spin-down states, respectively. }
		\label{fig1:Design}
	\end{figure}
		
    
	Here, we focus on a nonrelativistic perspective that excludes spin-orbit coupling effects. Consequently, the spin-space group $\left[R_s||R_l\right]$, along with time-reversal symmetry $T$, is appropriately suited for a symmetrical description, where $R_s$ and $R_l$ represent the spin and space groups, respectively. For stacking operations, we adopt the notation $\left\{G|\tau\right\}$, where $G$ denotes the point group operations and $\tau$ represents the sliding operations. 
	
	Significantly, from the perspective of stacking sliding, the condition $E\cap P = 0$ ensures that the stacking sliding operation $\left\{E|\tau\right\}$ is completely decoupled from $P$, thereby preserving the invariance of $\left[C_2||P\right]$.  Recalling previous predictions \cite{huang2025light,zhu2025floquet}, here $\left[C_2||P\right]$ serves as the key node for light-generated odd-parity altermagnets, independent of stacking configurations or light incidence directions. This feature highlights the stacking-sliding invariance of light-induced altermagnetism, which will be discussed concretely in the following sections.

	\begin{figure}
		\centering
		\includegraphics[width=1\linewidth]{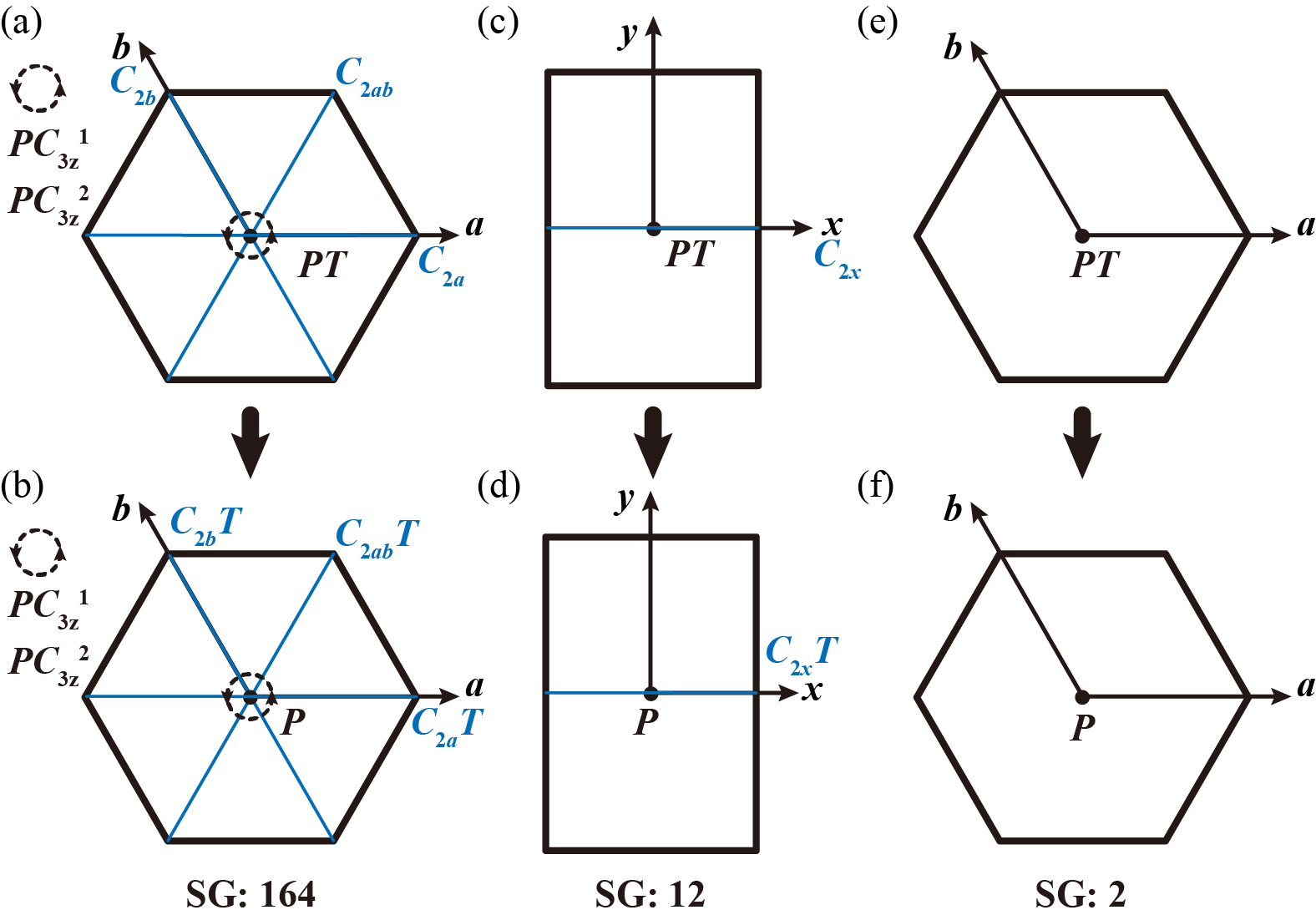}
		\caption{\textbf{Illustrations of the spin-space group symmetry operations that connect the two atomic layers of Mn. } Panels (a) and (b) show the space group No. 164 for the ground state and light-engineered state, with the latter having the irradiation direction along the $z$-axis. Blue solid lines represent rotation symmetry operations, while the dashed circle at the center indicates the rotation operators containing $C_{3z}$. Panels (c) and (d), as well as (e) and (f), are similar to (a) and (b), corresponding to space groups No. 12 and No. 2, respectively. The hexagonal and rectangular frames outline the shape and extent of one unit cell.}
		\label{fig2:Structure}
	\end{figure}

	\begin{table*}
		\caption{\label{tab1:SGs}Space groups of forward stacking BL-MBT resulting from various stacking sliding operations.}
		\begin{ruledtabular}
			\begin{tabular}{ccc}
				\multicolumn{1}{c}{Space Groups} &
				\multicolumn{1}{c}{Notations} &
				\multicolumn{1}{c}{Representative stacking operators}  \\ \hline
				\addlinespace[2pt]
				\multirow{1}{*}{No. 164} & \multirow{1}{*}{$P\bar{3}m1$}  & $\left\{E|\left(0,0\right)\right\}$, $\left\{E|\left(\frac{1}{3},\frac{2}{3}\right)\right\}$, $\left\{E|\left(\frac{2}{3},\frac{1}{3}\right)\right\} $   \\[2.7pt] 
				\multirow{1}{*}{No. 12} & \multirow{1}{*}{$C2/m$}  & $\left\{E|\left(\frac{1}{6},\frac{5}{6}\right)\right\}$, $\left\{E|\left(\frac{1}{2},0\right)\right\}$, $\left\{E|\left(\frac{1}{2},\frac{1}{2}\right)\right\}$, $\left\{E|\left(0,\frac{1}{2}\right)\right\}$, $\left\{E|\left(\frac{5}{6},\frac{1}{6}\right)\right\}$, $\left\{E|\left(\frac{1}{6},\frac{1}{3}\right)\right\}$, $\left\{E|\left(\frac{1}{3},\frac{1}{6}\right)\right\}$, etc. \\[2.7pt] 
				
				\multirow{1}{*}{No. 2} & \multirow{1}{*}{$P\bar{1}$}  & $\left\{E|\left(\frac{1}{6},0\right)\right\}$, $\left\{E|\left(0,\frac{1}{6}\right)\right\}$, $\left\{E|\left(\frac{1}{3},0\right)\right\}$, $\left\{E|\left(0,\frac{1}{3}\right)\right\}$,   $\left\{E|\left(\frac{1}{2},\frac{5}{6}\right)\right\}$, $\left\{E|\left(\frac{1}{2},\frac{1}{6}\right)\right\}$, $\left\{E|\left(\frac{1}{3},\frac{1}{3}\right)\right\}$, etc.    \\ [2.7pt]
			\end{tabular}
		\end{ruledtabular}
	\end{table*}

	\begin{table*}
		\caption{\label{tab2:forward} CPL-induced altermagnets or other magnetism in the forward stacking of BL-MBT. "CPL-$x$", "CPL-$y$", and "CPL-$z$" refer to CPL illumination along the $x$, $y$, and $z$ axes, respectively, while "Others" denotes incidence directions that do not align with any of these three axes.}
		\begin{ruledtabular}
			\begin{tabular}{cccccc}
				\multicolumn{1}{c}{Space Groups} &
				\multicolumn{1}{c}{Ground State} &
				\multicolumn{1}{c}{CPL-$x$} &
				\multicolumn{1}{c}{CPL-$y$} &
				\multicolumn{1}{c}{CPL-$z$} &
				\multicolumn{1}{c}{Others}
				  \\ \hline
				\addlinespace[1.7pt]
				\multirow{1}{*}{No. 164} & \multirow{1}{*}{AFM}  & $p$-wave &  $p$-wave  & $f$-wave  & $p$-wave  \\[1.7pt] 
				\multirow{1}{*}{No. 12} & \multirow{1}{*}{AFM}  & $p$-wave  &  $p$-wave  & $p$-wave & $p$-wave   \\[1.7pt]   \multirow{1}{*}{No. 2} & \multirow{1}{*}{AFM}  & $p$-wave  &  $p$-wave & $p$-wave & $p$-wave   \\[1.7pt]
			\end{tabular}
		\end{ruledtabular}
	\end{table*}

	\begin{figure*}
		\centering
		\includegraphics[width=0.95\linewidth]{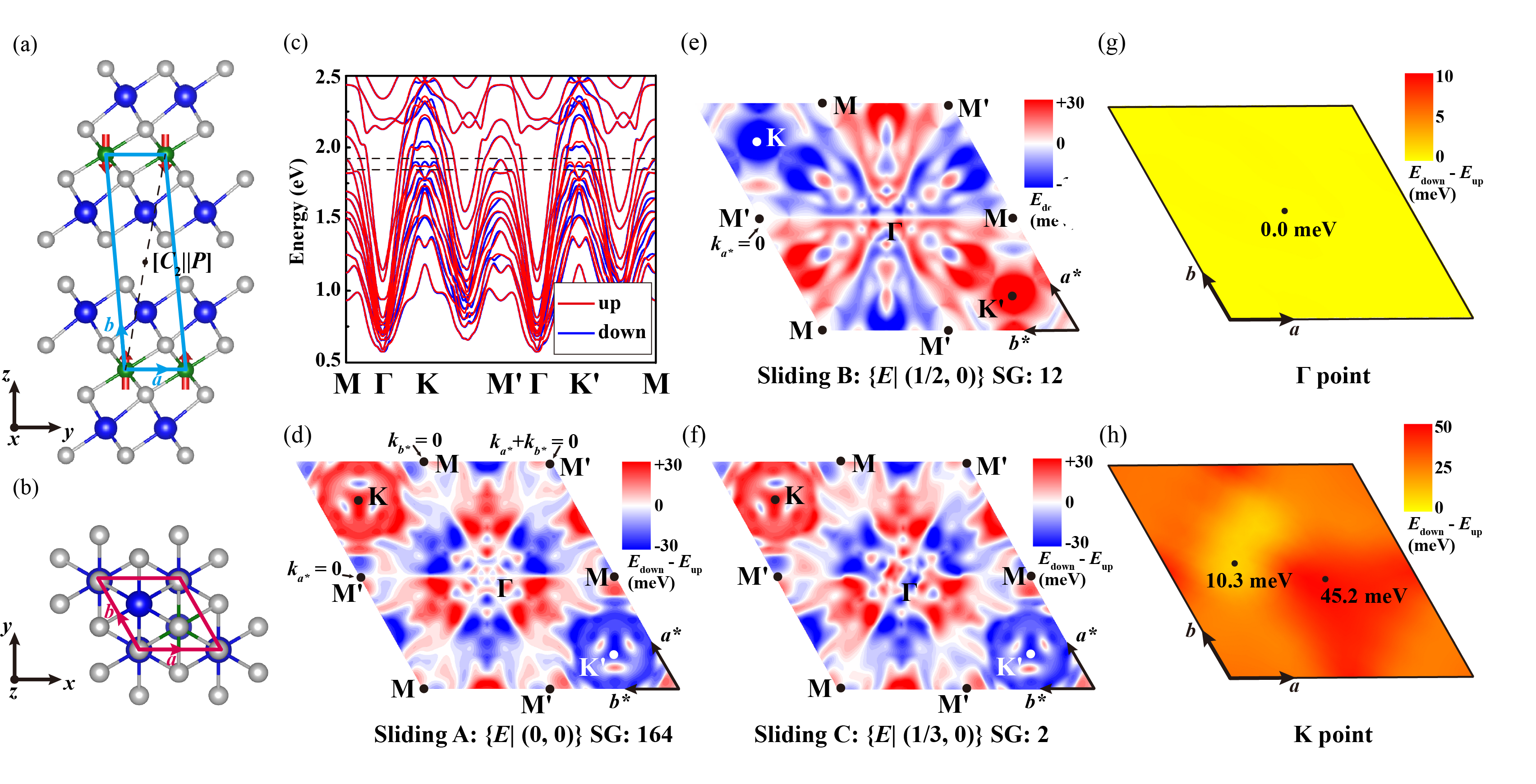}
		\caption{\textbf{BL-MBT: atomic structures, band structures along high-symmetry lines, and spin-splitting distributions for three stacking sliding positions, with light frequency set to $\hbar\omega = 9.0$ eV and incident direction along the $z$-axis.} Panel (a) shows a side view of BL-MBT, with olive, blue, and gray spheres representing Mn, Bi, and Te atoms, respectively; the light-blue rhombus frame indicates the unit cell. Panel (b) presents a top view, highlighting the stacking sliding range within the red rhombus frame. Panel (c) exhibits spin-resolved band structures along the high-symmetry lines $M$-$\Gamma$-$K$-$M'$-$\Gamma$-$K'$-$M$ under a light intensity of 0.3 Å$^{-1}$, with red and blue curves for spin-up and spin-down states, respectively. Panel (d) illustrates spin-splitting energy distributions across the entire BZ, with a color gradient from white to red (positive values) and blue (negative values) representing $E_{\mathrm{down}} - E_{\mathrm{up}}$. The selected occupancy number of the Wannier bands is 83, corresponding to panels (d) and (e), which are enclosed by a black dashed frame in panel (c). Panels (c) and (d) refer to the normal stacking order $\left\{E\middle|\left(0,0\right)\right\}$; panels (e) and (f) are similar for stacking sliding positions $\left\{E\middle|\left(\frac{1}{2},0\right)\right\}$ and $\left\{E\middle|\left(\frac{1}{3},0\right)\right\}$, respectively. Panels (g) and (h) show spin-splitting energies at the $\Gamma$ and $K$ points across the sliding zone, with a color gradient from yellow to red for $E_{\mathrm{down}} - E_{\mathrm{up}}$, with light incident along the $z$-axis.}
		\label{fig3:LDOSs}
	\end{figure*}

	Notably, A-AFM bilayers with $\left[C_2||P\right]$ symmetry can be formed from any arbitrary FM monolayer, as long as the interlayer coupling is AFM, driven by the $d$-electron counting rule \cite{xiao2020electron,li2020tunable,tang2023intrinsic}. The occupation numbers of $d$-electrons in the two magnetic atoms dictate the resulting magnetic couplings. Figure \ref{fig1:Design}(a) illustrates a straightforward method towards this. First, identify an operation center located just above the FM monolayer. Next, duplicate this monolayer after applying a $C_{2y}$ rotation, which transforms the coordinates $\left(x,\ y,\ z\right)$ to $\left(-x,\ y,\ -z\right)$, followed by a mirror operation with respect to the $x$-$z$ plane (denoted as $M_y$), resulting in the final coordinates of $\left(-x,\ -y,\ -z\right)$. This sequence of operations effectively achieves inversion symmetry, as depicted in the right panel of Fig. \ref{fig1:Design}(a). Thus, we propose a straightforward and universal method for creating a Floquet altermagnet from a FM monolayer, independent of stacking-sliding and CPL incident angle constraints, provided the interlayer coupling is AFM. Figure \ref{fig1:Design}(b) conspicuously illustrates this innovation in triggering $f$- or $p$-wave altermagnets from A-AFM bilayers. Candidate materials include bilayer MnBr$_2$, CoBr$_2$, NiBr$_2$, FeBr$_2$, MnBi$_2$Te$_4$ and LiFeSe, among others \cite{xiao2020electron,li2019intrinsic,li2020tunable,li2020high}.


	We use BL-MBT as a representative platform to illustrate the principles discussed. MBT is a magnetic topological insulator with FM coupling within monolayers and AFM coupling between layers in the bilayer form, following an "$ABCABC$…" stacking sequence \cite{li2019intrinsic,otrokov2019prediction,otrokov2019unique,zhang2019topological,li2020tunable,zhang2020large,tang2023intrinsic,deng2020quantum,gong2019experimental,liu2020robust,ge2020high,bai2024quantized,li2024multimechanism,li2025universally}. BL-MBT crystallizes in a trigonal lattice of space group No. 164 ($P\bar{3}m1$), maintaining inversion symmetry $P$. The two atomic layers of Mn, with opposite spins, are linked by symmetries $\left[C_2||P\right]$ and $\left[C_2||T\right]$, ensuring spin degeneracy across the BZ. While stacking-sliding preserves $\left[C_2||P\right]$, it reduces overall symmetry. As shown in Table \ref{tab1:SGs}, forward stacking yields three space groups: No. 164, No. 12 ($C2/m$), and No. 2 ($P\bar{1}$). Figure S1 \cite{supplementary} displays the lattice vectors and the first BZ, while Table \ref{tab2:forward} details the magnetic configurations at ground states and under CPL irradiation along the $x$, $y$, and $z$ axes, as well as other directions, all exhibiting odd-parity altermagnetism.
	
	Figure \ref{fig2:Structure} schematically illustrates the CPL-induced symmetry reduction for these three space groups under out-of-plane irradiation. For space group No. 164, vertical CPL irradiation breaks $\left[C_2||T\right]$ and mirror symmetries while preserving the rotational operators, including $\left[E||C_{2a}T\right]$, $\left[E||C_{2b}T\right]$ and $\left[E||C_{2ab}T\right]$, which together protect an $f$-wave altermagnetic state. By contrast, the absence of these combining operators in space groups No. 12 and No. 2 leads to $p$-wave altermagnet: in the former case, a spin degenerate line survives due to $\left[E||C_{2x}T\right]$	[Fig. \ref{fig2:Structure}(d)], whereas in the latter only a single degenerate point at $\Gamma$ remains, projected  by $\left[C_2||P\right]$
	[Fig. \ref{fig2:Structure}(f)]. Likewise, for space groups with reversed stacking that allow for altermagnetism, we present the CPL-induced symmetry degradation in Fig. S2 \cite{supplementary}.
	
	We focus next on the electronic structure of forward-stacked BL-MBT. Figure \ref{fig3:LDOSs}(a-b) shows the side and top views of the bilayer and the representative sliding configuration.  For simplicity,  CPL is taken to propagate along the $z$-axis. In the normal stacking configuration (sliding A, space group No. 164), the remaining spin-space symmetries (see details in Sec. I of Supplementary Materials \cite{supplementary}) generate three symmetry related spin-degenerate lines at $k_{a^{*}}=0$, $k_{b^{*}}=0$ and $k_{a^{*}}+k_{b^{*}}=0$. Figure \ref{fig3:LDOSs}(c) displays the spin-resolved band structures along the high-symmetry path $M$-$\Gamma$-$K$-$M'$-$\Gamma$-$K'$-$M$ [as illustrated in the first BZ in Fig. S1(b)] under CPL irradiation with intensity $A$ = 0.3 Å$^{-1}$, and frequency $\hbar\omega$ = 9.0 eV, chosen to avoid  hybridization between Floquet replicas. While no spin splitting appears in the absence of light 
    (Fig. S3 \cite{supplementary}) , CPL induces opposite spin splitting on symmetry-related momentum lines along  $M$-$\Gamma$-$K$-$M'$ and $M'$-$\Gamma$-$K'$-$M$, providing direct evidence of odd-parity altermagnetism. The distribution of spin-splitting energies over the entire BZ [see Fig. \ref{fig3:LDOSs}(d)] further confirms the $f$-wave character, with three spin-degenerate lines arranged at 60$^{\circ}$ angles to each other, with  dominant contributions from Mn 3$d$ orbitals (see Fig. S4 \cite{supplementary}).
    
	Stacking sliding continuously lowers the symmetry and dirves a transition from $f$-wave to $p$- wave altermagnetism. For example, the sliding configuration $\left\{E|\left(\frac{1}{2},0\right)\right\}$ (sliding B, space groups No. 12, $C2/m$) retains only $\left[C_2||P\right]$ and $\left[E||C_{2x}T\right]$, protecting a single spin-degenerate lin at $k_{a^{*}}=0$, [Fig. \ref{fig3:LDOSs}(e)].Despite this degradation, the opposite spin splitting remains evident along the corresponding reciprocal lines [Figs. S5(a)]. Further sliding to $\left\{E|\left(\frac{1}{3},0\right)\right\}$ (sliding C, group No.2, $P\bar{1}$) leaves only 
    $\left[C_2||P\right]$, enforcing the relation $\varepsilon(\textbf{\textit{k}},\ s)=\varepsilon(-\textbf{\textit{k}},\ -s)$. As a result, spin degeneracy solely at the $\Gamma$ point ($k_{a^{*}}=k_{b^{*}}=0$) with an antisymmetric manner around [Figs. S5(b) and \ref{fig3:LDOSs}(f)].
    
    \begin{table*}
    	\caption{\label{tab3:reversed} CPL-induced altermagnets or other magnetism in the reversed stacking of BL-MBT. }
    	\begin{ruledtabular}
    		\begin{tabular}{cccccc}
    			\multicolumn{1}{c}{Space Groups} &
    			\multicolumn{1}{c}{Ground State} &
    			\multicolumn{1}{c}{CPL-$x$} &
    			\multicolumn{1}{c}{CPL-$y$} &
    			\multicolumn{1}{c}{CPL-$z$} &
    			\multicolumn{1}{c}{Others}
    			\\ \hline
    			\addlinespace[1.7pt]
    			\multirow{1}{*}{No. 187} & \multirow{1}{*}{AFM}  & $p$-wave &  $p$-wave  & AFM  & $p$-wave or CFiM  \\[1.7pt] 
    			\multirow{1}{*}{No. 156} & \multirow{1}{*}{CFiM}  & CFiM  &  CFiM  & CFiM & CFiM  \\[1.7pt]   \multirow{1}{*}{No. 39} & \multirow{1}{*}{AFM}  & $p$-wave  &  $p$-wave & AFM & $p$-wave or CFiM   \\[1.7pt]
    			\multirow{1}{*}{No. 8} & \multirow{1}{*}{CFiM}  & CFiM  &  CFiM  & CFiM & CFiM  \\[1.7pt]
    			\multirow{1}{*}{No. 5} & \multirow{1}{*}{$d$-wave}  & $p$-wave  &  $p$-wave & $p$-wave & CFiM   \\[1.7pt]
    			\multirow{1}{*}{No. 1} & \multirow{1}{*}{CFiM}  & CFiM  &  CFiM  & CFiM & CFiM  \\[1.7pt]
    		\end{tabular}
    	\end{ruledtabular}
    \end{table*}	    
%
	
	\begin{figure}
		\centering
		\includegraphics[width=1\linewidth]{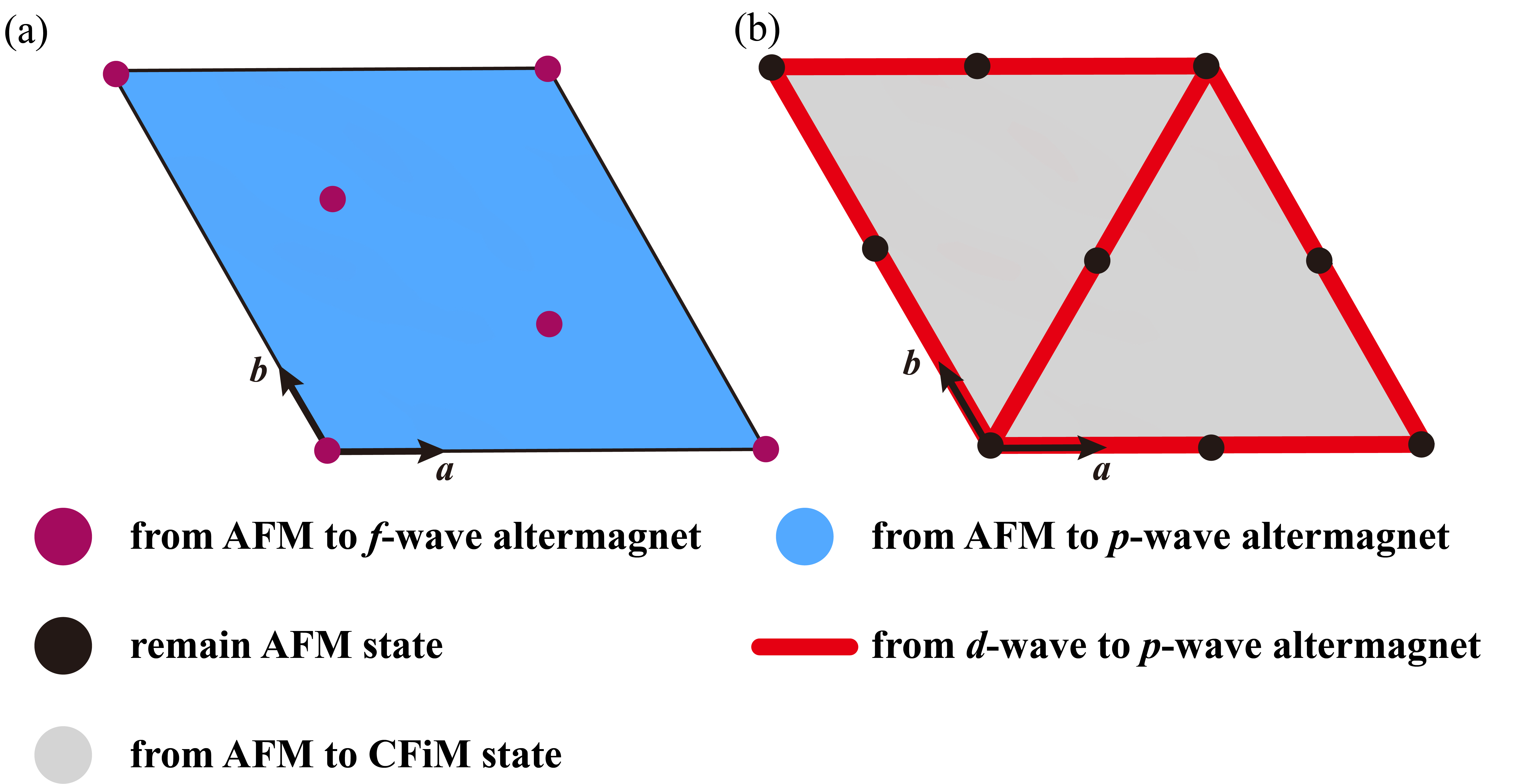}
		\caption{Phase diagrams in (a) forward stacking and (b) reverse stacking of bilayer BL-MBT under the \textbf{out-of-plane} irradiation.  Each phase diagram represents the sliding positions of various magnetic configurations within a single sliding cell.}
		\label{fig4:Contours}
	\end{figure}
	
	Importantly, the odd-parity altermagnetic state is invariant under stacking sliding.  Figures \ref{fig3:LDOSs} (g) and \ref{fig3:LDOSs}(h) show the spin-splitting energies at the $\Gamma$ and $K$ points across the entire sliding cell. While spin degeneracy is strictly preserved at $\Gamma$, finite spin splitting persists at $K$ for all sliding positions, with magnitudes ranging from   10.3 to 45.2 meV. This behavior directly demonstrates the stacking-sliding invariance of Floquet-induced altermagnetism in forward-stacking BL-MBT. The corresponding phase diagram is summarized in Fig. \ref{fig4:Contours}(a), where CPL produces three points of $f$-wave altermagnetism while maintaining the others in the $p$-wave state. 
    
	As shown in Table \ref{tab2:forward}, when the incident CPL direction deviates from the $z$-axis, the combined symmetry operator of $\left[E||C_{2a}T\right]$, $\left[E||C_{2b}T\right]$ and $\left[E||C_{2ab}T\right]$ is broken, resulting in the loss of $f$-wave characteristics. However, the $p$-wave altermagnetic states remain robust, protected by $\left[C_2||P\right]$. For the three  stacking configurations, the surviving symmetry operations in different directions are summarized in Tables S1-S3  \cite{supplementary}, and the representative band structures for sliding A with the incident angles along $x$, $y$ axes and other direction are shown in Fig. S6 \cite{supplementary}. Furthermore, a detailed examination of the theoretical analysis on forward stacking of BL-MBT is presented in Sec. I of the Supplementary Materials \cite{supplementary}.
	
	We then contrast this behavior with reversed stacking of BL-MBT [Fig. S6(a) \cite{supplementary}], where $\left[C_2||P\right]$ is replaced by $\left[C_2||M_z\right]$. The ground-state remains AFM and the system crystallizes in space group No. 187 ($P\bar{6}m2$), which breaks the inversion symmetry. However, for $x$-axis incidence, the symmetry $\left[C_2||M_z\right]$ is broken, allowing a $p$-wave altermagnetic state. For a generic CPL incidence direction, except for the cases when light is incident along the $x-y$ plane which also give rise to a $p$-wave altermagnet, no spin-space symmetry remains to relate the two Mn sites with opposite spins. Instead, the system transitions into a compensated ferrimagnetic (CFiM) state \cite{liu2025robust-CFiM}, exhibiting spin splitting throughout the entire BZ. Consequently, without the protection of $\left[C_2||P\right]$, the altermagnetic response is no longer invariant with respect to the CPL incident direction. Further details are provided in Table S4 and Fig. S7 in Secs. I and IV of the Supplementary Materials \cite{supplementary}.
	
	Additionally, reversed stacking with in-plane sliding produces five more space groups: No. 156 ($P3m1$), No. 39 ($Aem2$), No. 8 ($Cm$), No. 5 ($C2$), and No. 1 ($P1$). Table \ref{tab3:reversed} concludes the corresponding magnetic configurations at ground state and under CPL illumination from four incident directions. Among them, only No. 39 ($Aem2$) and No. 5 ($C2$) support a $p$-wave altermagnetism, and only for specific illumination directions. Remarkably, space group No. 5 exhibits a ground-state $d$-wave altermagnet that transitions to a $p$-wave altermagnet under out-of-plane CPL irradiation, as shown in Fig. S8 \cite{supplementary}. This serves as a concrete example of light-induced parity conversion of altermagnetism from even to odd. For the remaining three space groups, the CFiM state persists across the illumination conditions considered. The phase diagram for reversed stacking within one sliding cell is summarized in Fig. \ref{fig4:Contours}(b). The $d$-$p$ conversion occurs at the positions indicated by the red lines, while $x$-axis irradiation restores altermagnetism from AFM at specific sliding configurations (black points in Fig. \ref{fig4:Contours}(b)). Detailed analyses are provided in Tables S5-S7 in Sec. I of the Supplementary Materials \cite{supplementary}.
	
	In summary, we establish a symmetry based route to engineer a controllable Floquet altermagnetism in A-type AFM bilayers trough stacking sliding and light light incidence geometry. Using spin-space group analysis, we show that when inversion symmetry is present, CPL/EPL driving can generate odd-parity altermagnetism that is robust against stacking sliding and, in most cases, illumination direction. We further proposed a general construction of inversion-symmetric A-AFM bilayer from 
    arbitrary ferromagnetic monolayers with antiferromagnetic interlayer coupling, guided by the $d$-electron counting rule. Taking forward-stacking BL-MBT as a representative example, we demonstrate that out-of-plane CPL stablizes an $f$-wave altermagnet, while symmetry lowering by sliding or by changing the incident direction tunes the system into a $p$-wave state altermagnet. Crucially, we also analyze the reverse-stacking case, where inversion symmetry is absent and the altermagnetic response becomes strongly dependent on illumination direction, competing with compensated ferrimagnetism. The resulting phase diagrams over the full sliding cell, for both forward and reverse stacking, provide a practical blueprint for symmetry-guided design and tunable control of altermagnets in realistic bilayer materials.
	
	\textit{Note added.} We become aware of a recent study on Floquet engineering of the transition from A-type AFM to altermagnets in two-dimensional bilayers found that only bilayers with initial centrosymmetric point groups \(D_{3d}\) or \(C_{2h}\) can undergo this phase transition \cite{pan2025floquet}. Advanced than this work, our study demonstrates the invariance of Floquet altermagnetism against stacking sliding and light incidence directions, provided inversion symmetry is maintained, even with $M_z$ present, except under incidence along $z$-axis. Besides, we propose Floquet-induced altermagnetic parity conversion from even to odd for specific stacking orders of BL-MBT. These findings offer a fundamental blueprint for designing and transforming interleaved magnetism across a broader range of materials.

	\begin{acknowledgments}
		We thank Prof. Huisheng Zhang, Dr. Xiyu Hong, Dr. Chaoxi Cui, Ling Bai, Yongli Xie, Jingjing Cao and Wenlu Zhang for helpful discussions, and thank Asso. Prof. Huixia Fu, Dr. Hang Liu for technical supports. The numerical calculations have been done on the supercomputing system in the Huairou Materials Genome Platform. This work is supported by National Natural Science Foundation of China (Grant Nos. 12025407, 12450401, 12104072 and 12304536), National Key Research and Development Program of China (Grant No. 2024YFA1207800 and No. 2021YFA1400201), and Chinese Academy of Sciences (Grant Nos. YSBR-047 and No. XDB33030100).  
		
		Zhe Li and Lijuan Li contributed equally to this work.
	\end{acknowledgments}

	\newpage
	\nocite{*}

%

\end{document}